\documentclass{amsart}
\usepackage{amssymb,latexsym,amsmath}
\usepackage{setspace}
\usepackage{graphics}
\usepackage{psfrag}
\usepackage{graphicx}

\input epsf.sty

\theoremstyle{plain}
\newtheorem{theorem}{Theorem}[section]

\newtheorem{lemma}[theorem]{Lemma}
\newtheorem{proposition}[theorem]{Proposition}

\numberwithin{equation}{section}

\font\cmcsc=cmcsc10 at 8pt

\begin{document}

\title[Call Derivatives]{The derivatives of Asian call option prices}
\author{ Jungmin Choi }
\address{Mathematics Department\\
         Florida State University\\
         Tallahassee, FL 32306}
\email{choi@math.fsu.edu}

\author{ Kyounghee Kim }
\address{Mathematics Department\\
         Florida State University\\
         Tallahassee, FL 32306}
\email{kim@math.fsu.edu}

\maketitle

\begin{abstract}
The distribution of a time integral of geometric Brownian motion is not well understood. To  price an Asian option and to obtain measures of its dependence on the parameters of time, strike price, and underlying market price, it is essential to have the distribution of time integral of geometric Brownian motion and it is also required to have a way to manipulate its distribution. We present  integral forms for key quantities in the price of Asian option and its derivatives ({\it{delta, gamma,theta, and vega}}).  For example for any $a>0$ $\mathbb{E} \left[ (A_t -a)^+\right] = t -a +  a^{2} \,\mathbb{E} \left[ (a+A_t)^{-1} \exp ( \frac{ 2M_t}{a+ A_t} - \frac{2}{a} ) \right]$, where $A_t = \int^t_0 \exp (B_s -s/2)\, ds$ and $M_t =\exp (B_t -t/2).$ \hfill\break
\phantom{as}\hfill\break
\noindent {\cmcsc Keywords}:\ Asian option, Derivatives of option prices, Geometric Brownian Motion, Time integral.\hfill\break
\noindent {\cmcsc Subject Classification}:\   Primary 91B28, 60J65\ \      Secondary 60G99.
\end{abstract}

\vspace{4ex}

\section{Introduction}

The payoff of an Asian option depends on the (geometric or arithmetic) average of prices of a given risky asset over the pre-specified time interval. Under the Black-Scholes framework, one assumes that the price process $\{S_t, t \ge 0\}$ of the risky asset follows $$dS_t = \mu S_t dt + \sigma S_t d B_t, \qquad S_0 >0 $$ where $\mu$ and $\sigma$ are given constants and $\{B_t, t\ge 0\}$ is a standard one dimensional Brownian motion. In this setting, it is easy to understand the geometric average; if $0 \le t_1 < t_2$ then in distribution $$ \sqrt{S_{t_1}\cdot S_{t_2}} = S_0 \exp \left[\ ( \sigma \sqrt{t_2+ 3 t_1})\cdot \mathcal{N} + (\mu- \frac{\sigma^2}{2}) (t_1+t_2)\ \right] $$ where $\mathcal{N}$ is a standard normal random variable. On the other hand, the distribution of an arithmetic average process is not well understood. A continuous version of the arithmetic average is  a time integral of a price process. Using the Inverse Lapalce Transformation, Yor \cite{Y} proved many interesting identities related to the distribution of geometric Brownian motion, which gives us deeper understanding of functions of geometric Brownian motion and useful information about their time integrals. More detailed research for the relation between the time integral and an Asian option was considered in \cite{GY}. Using the joint density  of $( \int^t_0 \exp(B_s) dW_s, \exp (B_t) ) $ where $B_t, W_t$ are independent Brownian motions (given in \cite{B}) the moment generating function of the time integral process was computed in \cite{K} . The method of changing measures was considered to analyze the properties of the time integral process. (See \cite{GK} \cite{MY1} \cite{MY2} \cite{MY3}.) In \cite{D} the very useful time reversing property is used to analyze the time integral process. Dufresne also provided a certain form for  the density function of the time integral of geometric Brownian motion. However the author pointed out the difficulties to use his formula in practice especially when the time integral is over the short time period due to the slow convergence rate.   

The payoff of an European style fixed strike Asian option is given by a function of the time integral of the price of the risky asset $S_t$ $$(\frac{1}{|I|} \int_I \, S_t \, dt- \kappa)^+$$ where $\kappa$ is a fixed strike price and $I$ is the pre-specified time interval with the length $|I|$. Under the risk neutral measure $\mathbb{Q}$, we may set the price process $S_t$ given by an SDE, $dS_t = \sigma S_t dB_t$ where $\sigma$ is a constant depending on the risky asset and $B_t$ is a $\mathbb{Q}$-brownian motion. Without loss of generality, we may also assume that the time interval $I=[0, \tau]$ for some $\tau>0$. It follows that the price of European style Asian option is given by  
$$ e^{-r\tau} E_{\mathbb{Q}}\left[(\frac{1}{\tau} \int_0^\tau S_0 \exp\{\sigma B_t - \sigma^2t/2\} \,dt - \kappa)^+\right ].$$
Since $\sigma B_t = B_{\sigma^2 t}$ in law, we can rewrite the above quantity as follows
\begin{equation}\label{E:asian}\frac{S_0}{\tau \sigma^2}e^{-r\tau} E_{\mathbb{Q}}\left[( \int_0^{\sigma^2\tau}  e^{ B_t - t/2}\, dt - \frac{\sigma^2\kappa\tau}{S_0})^+\right ].\end{equation} Note that the initial price of the risky asset is not appeared in the time integral. To obtain the the price of Asian option and the derivatives with respect to the asset price we need to understand the following quantities 
\begin{equation}\label{E:quantities} E\left[ (A_t - a)^+\right], \ \ \frac{d}{da} E\left[ (A_t - a)^+\right], \ \ {\rm and\ \ } \frac{d^2}{da^2} E\left[ (A_t - a)^+\right]\end{equation} where $A_t =  \int^t_0 \exp (B_s -s/2)\, ds$ and $a >0$. In this paper we show that the quantities in (\ref{E:quantities})  can be expressed in terms of the expected values of functions of exponential Brownian motion. We believe that these expressions would provide the alternative approach to simulate the Asian option price and its greeks. The simulation results for Asian option price and its greeks were considered by several authors.  (For example see \cite{A}, \cite{BG}, \cite{CSW}, \cite{J}.) 

Other types of Asian options are also considered. When the strike price depends on the average price, it is called the floating-strike Asian option. (See for example \cite{HHSW} \cite{HW}.) In \cite{HW}, Henderson and Wojakowski show the very useful symmetries between fixed-strike and floating strike Asian options. They showed that at the starting point of the averaging period there exists an equivalent relation between the floating-strike Asian option and the fixed strike Asian option. However, once the averaging period has begun, the floating strike Asian option can not be re-expressed as a fixed strike option. 

In section 2 we discuss the relation between the time integral and the exponential Brownian motion. We summarize the result in \cite{GK} and present the key proposition. In section 3 we discuss the price of an Asian option and its derivatives (delta, gamma, theta, and vega).

\bigskip

\section{The time integral of exponential Brownian motion.} 

For $\mu \in \mathbb{R}_{\ge 0}$ let us denote $A_t^{(\mu)}$ a time integral of an exponential Brownian motion with drift $\mu$ $$A_t^{(\mu)} := \int^t_0 \exp [B_s + (\mu-1/2) s ] \, ds. $$ When $\mu =0$ we simply use $A_t$ without a superscript. For $ t \ge 0$ and $y >0$ we set $M_t := \exp (B_t-t/2)$ and  
$$R_t:= - \frac{M_t}{y^{-1}-1/2 \, A_t}\,=\, 2\frac{d}{dt}\log(1-\frac{y}{2}A_t) .$$ It is not hard to see that the process $R_t$ satisfies the following SDE 
\begin{equation}
dR_t =R_tdB_t - \frac{1}{2} R_t^2 \,dt,\ \ \ \ R_0=-y
\end{equation}
up to an explosion time $\tau_\infty := \inf \{ t\ge 0: A_t = 2/y\}$. By considering a stopping time $\tau_n = \inf \{t : R_t \le -n\}$ and a Girsanov density  process $R_t 1_{\{t < \tau_n\}}$, we define a new measure $\mathbb{Q}$ under which $$\tilde B_t = B_t - \int^t_0 R_s 1_{\{s < \tau_n\}} \, ds= B_t - 2 \log(1-\frac{y}{2} A_{t \wedge \tau_n})$$ is a standard Brownian motion. Let us define $\tilde M = \exp (\tilde B_t -t/2)$ and $\tilde A_t = \int^t_0 \tilde M_s\, ds$. It follows that 
\begin{equation}\label{E:dic1}
\tilde M_t = \frac{M_t}{(1-\frac{y}{2}A_{t \wedge \tau_n})^2},\ \ \ 1+\frac{y}{2} \tilde A_{t \wedge \tau_n} = \frac{1}{1-\frac{y}{2}A_{t \wedge \tau_n}}\end{equation}
and \begin{equation}\label{E:dic2}R_{t\wedge \tau_n}= - \frac{\tilde M_{t \wedge \tau_n}}{y^{-1}+ \frac{1}{2} \tilde A_{t \wedge \tau_n}}.\end{equation}
The following proposition is a simple modification of the proposition 3.3 in \cite{GK}.

\begin{proposition}\label{T:keylemma}
If $f(x,z)$ is a Borel measurable function and $y>0$ then
\begin{equation}\label{E:key}
\begin{aligned}
E&\left[ f(M_t, A_t); A_t < \frac{2}{y} \right] \\&= e^{-y} E\left[ f(\frac{M_t}{(1+\frac{y}{2} A_t)^2}, \frac{A_t}{1+\frac{y}{2}A_t}) \exp \left( \frac{M_t}{y^{-1}+\frac{1}{2}A_t}\right) \right].
\end{aligned}
\end{equation} 
\end{proposition}

\begin{proof} The proof of the proposition is essentially same as the proof of the proposition 3.3 in \cite{GK}, so we give a brief sketch. (For the detailed proof we refer \cite{GK}.) For fixed $n$ and $y>0$ it is not hard to see that 
$$ \exp(R_t+y)1_{\{\tau_n>t\}} = \exp ( \int^{t \wedge\tau_n}_0 R_s dB_s - \frac{1}{2} \int^{t \wedge \tau_n}_0R_s^2\, ds)$$ satisfies a Novikov condition. Thus we have 
$$E\left[f(M_t, A_t); \tau_n >t\right] = E_{\mathbb{Q}}\left[ f(M_t, A_t)\exp(-R_t-y)1_{\tau_n>t}\right].$$
Since the event $\{ \tau_n>t\}$ is the same as $\{ \max_{s \le t} \tilde M_s/(y^{-1}+ 1/2 \, \tilde A_s)<n\}$, by letting $n\to \infty$ the r.h.s. becomes $E_{\mathbb{Q}}[f(M_t,A_t) \exp (-R_t-y)]$. Use (\ref{E:dic1}) and (\ref{E:dic2}) we can rewrite the limit of the r.h.s. in terms of $\tilde M_t, \tilde A_t$ and $\tilde R_t$. Since we are only interested in the quantity, we remove the tilde and get (\ref{E:key}).  \end{proof}

Let us discuss certain of interesting choices of $f$ in Proposition \ref{T:keylemma}. Simple choices like a constant function or power functions allows us to have various relations between a simple martingale $M_t$ and its time integral. We consider the ones who directly related to the problem of pricing the European style Asian option and estimating the sensitivities. First we consider a constant function $f(x,z)\equiv 1$. Using a constant function, it is easy to see that the l.h.s. of (\ref{E:key}) is a probability distribution of the time integral of a geometric brownian motion. By setting $a= 2/y$ it follows that 
\begin{lemma}\label{L:distribution1}  
For a positive $a>0$, we have
\begin{equation*}
Pr[A_t < a]  = e^{-2/a} E \left[ \exp\left(\frac{M_t}{ a/2 +1/2\,A_t}\right)\right].
\end{equation*}
\end{lemma}

\vspace{1ex}

\noindent For $\nu \in \mathbb{R}$ we have $M_t^\nu \exp(\nu t/2-\nu^2t/2) = \exp (\nu B_t -\nu^2t/2)$ and thus \begin{equation}\label{E:drift} E[M_t^\nu \exp(\nu t/2-\nu^2t/2) ; A_t \le 2/y] = Pr[A_t^{(\nu)}\le 2/y].\end{equation}
By setting $f(x,z)=x^\nu$ we get
\begin{lemma}\label{L:distribution2} 
For a real number $\nu$ and a positive real $a>0$, we have
\begin{equation}
\begin{aligned}
Pr& [A_t^{(\nu)} \le a] \\&= a^{2 \nu} e^{-2/a} \mathbb{E} \left[ (a+ A_t^{(\nu)})^{-2\nu} \exp \left ( \frac{2 \exp ( B_t + (\nu-1/2) t)}{a+ A_t^{(\nu)}}  \right) \right].
\end{aligned}
\end{equation}
\end{lemma}
\vspace{1ex}

\centerline{\epsfysize=1.8in \epsfbox{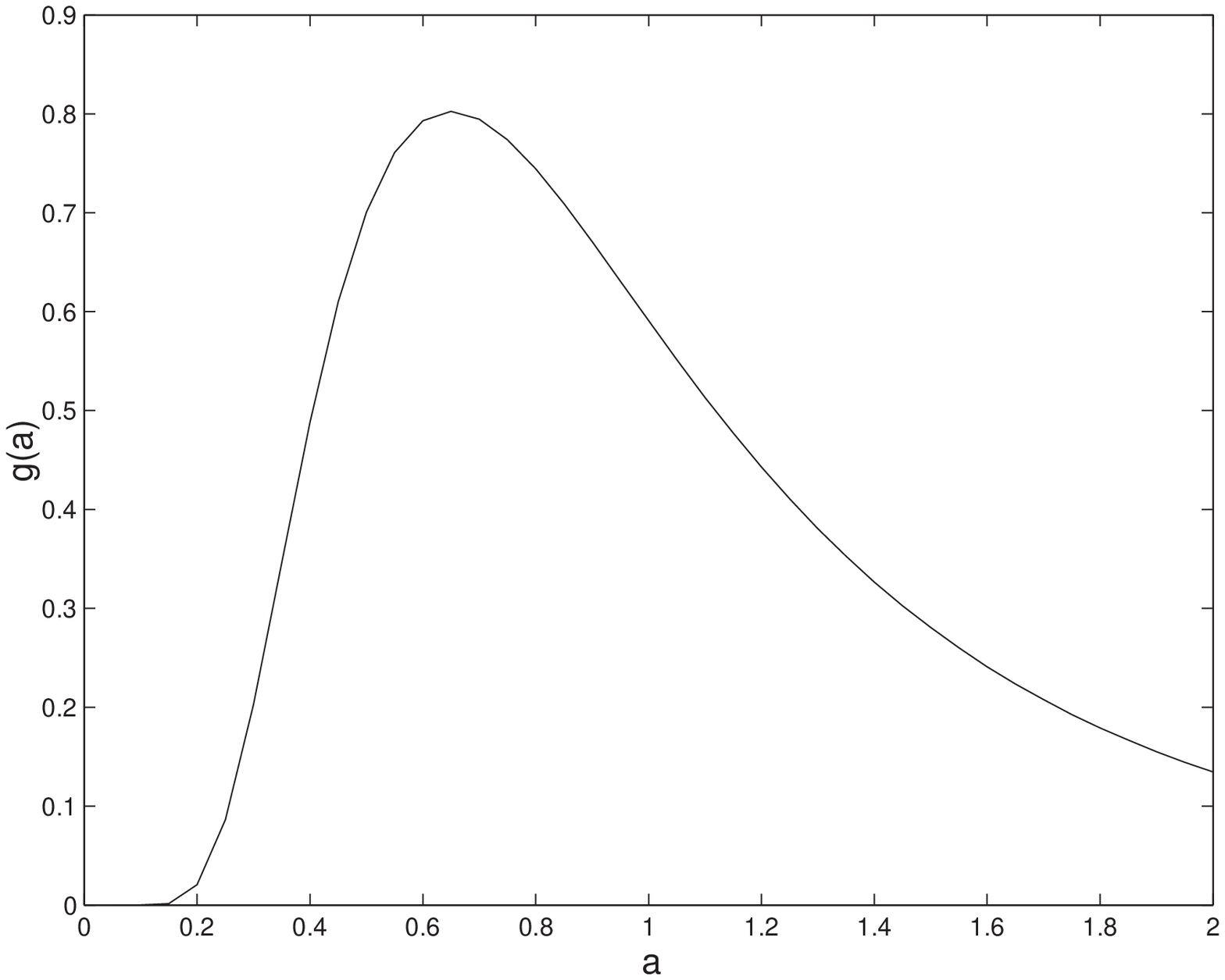}}
\centerline{Figure 2.1. The density function of $A_t$. }

\vspace{1ex}

Let $g_t$ denote the probability density function for $A_t$. It is known that the density function for $A_t$ is continuous and positive. Figure 2.1 is the result of the simulation ($10000$ simulation for each $a$) using the identity given in the following Lemma \ref{L:density}. The $n$-th moments of the time integral process $A_t$ can be computed with simple computation. However it is known that $A_t$ has a heavy tail probability and thus knowing all integer moments does not give a probability density function. In Figure 2.1 we can see the heavy tail probability.

\begin{lemma}\label{L:density}
For a real number $a>0$, the probability density function $g_t$ for $A_t$ satisfies 
$$g_t(a) = \frac{2}{a^2} Pr[A_t\le a] - \frac{2}{a^2} Pr[A_t^{(1)} \le a].$$
\end{lemma}
\begin{proof} From Lemma \ref{L:distribution1} we have 
\begin{equation*}
\begin{aligned}
g_t(a) &= \frac{d}{da} Pr[A_t <a]\\ & =\frac{2}{a^2} Pr[A_t\le a] - \frac{1}{2}e^{-2/a} E\left[\frac{M_t}{(a/2+1/2\, A_t)^2} \exp\left( \frac{M_t}{a/2+1/2\, A_t}\right)\right]
\end{aligned}
\end{equation*}
Since $M_t = \exp (B_t - t/2)$ we use $M_t$ as a Girsanov density function for the second term. Under the new measure $B_t - t/2$ is a standard Brownian motion and thus the second term is equal to the following quantity:
$$\frac{1}{2} e^{-2/a} E\left[ \frac{1}{(a/2+1/2\, A_t^{(1)})^2} \exp \left( \frac{2\exp(B_t + t/2)}{ a+A_t^{(1)}}\right) \right].$$
Comparing with Lemma \ref{L:distribution2}, we can see that the above quantity is the same as $2/a^2 Pr[A_t^{(1)} \le a]$. 
\end{proof}

\noindent Furthermore we have

\begin{proposition}\label{P:joint} For any $a>0$, $b>0$ and $t>0$ we  have
$$ Pr[M_t <b, A_t <a] = e^{-2/a} \mathbb{E} \left[ \exp \left( \frac{2 M_t}{a+A_t}\right) \, ;\, M_t \le b (1+ \frac{1}{a} A_t)^2\right].$$
\end{proposition}
\begin{proof} Consider $f(x,z) = 1\{x <b\}$ in equation (\ref{E:key}). By setting $a= 2/y$ we get an integration form for the joint probability density function for $M_t$ and $A_t$. \end{proof}
The simulation result of the above proposition \ref{P:joint} for $t=a$ is shown in the l.h.s. of the following figure and the joint distribution for $t=1$ is shown in the r.h.s. 

\vspace{1ex}

\centerline{\epsfysize=1.8in \epsfbox{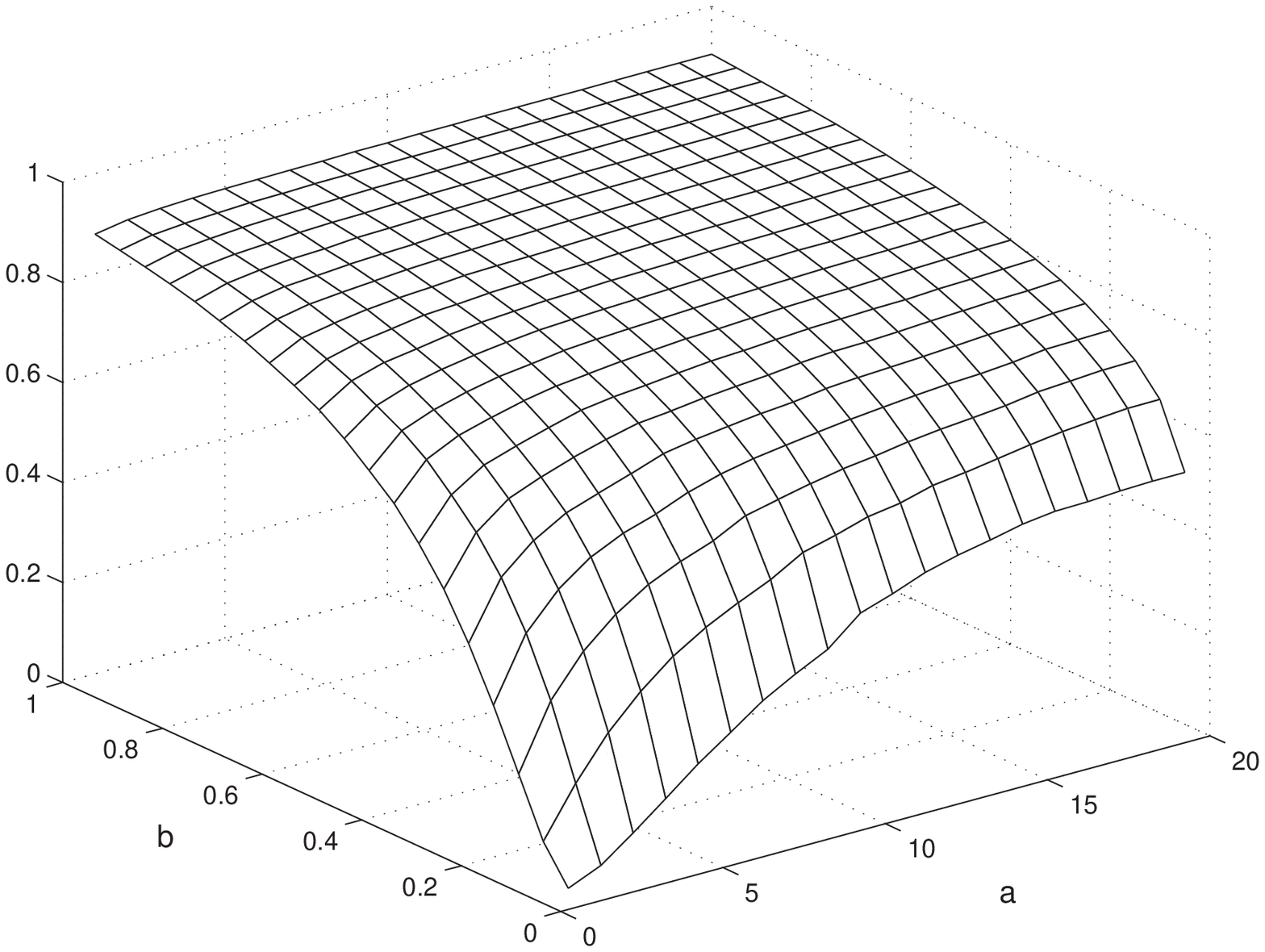}\quad \epsfysize=1.8in \epsfbox{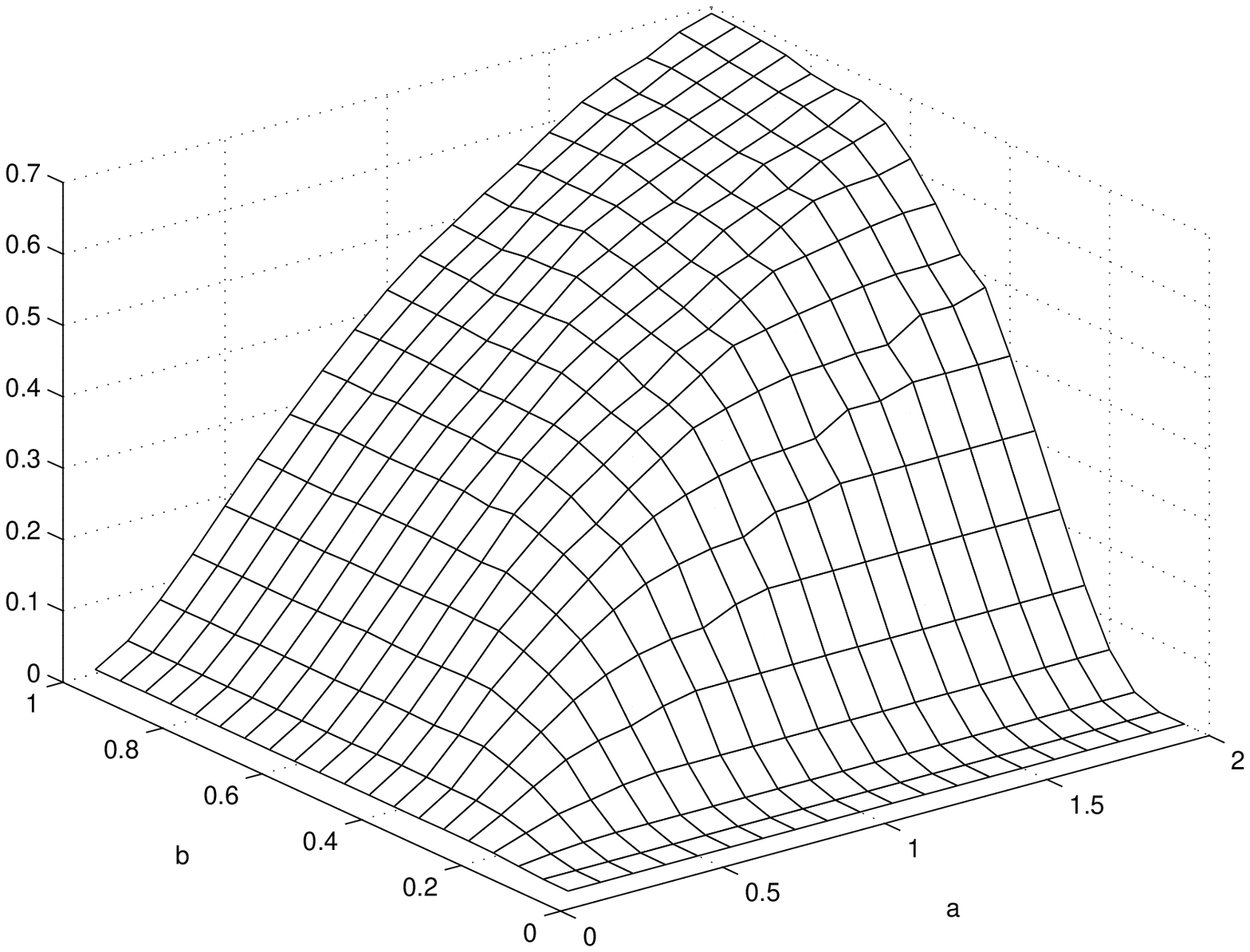}}
\centerline{Figure 2.2. Joint distribution of $M_t$ and $A_t$. }

\vspace{1ex}

The first quantity of interests to have the derivatives of Asian options is the expected value of the maximum function of the time integral subtracted by a constant. Using Proposition \ref{T:keylemma} we have the following result :

\begin{theorem}\label{T:moment} For any $a>0$ and $\nu\in \mathbf{R}$, we have 
$$\mathbb{E} \left[ (A_t^{(\nu)} -a)^+\right] = \frac{e^{\nu t} -1}{\nu} -a + e^{\nu t /2 - \nu^2 t/2 } a^{2 \nu +2} \,\mathbb{E} \left[ \frac{M_t^\nu}{(a+A_t)^{2 \nu+1} }\exp ( \frac{ 2M_t}{a+ A_t} - \frac{2}{a} ) \right].$$
In particular if $\nu = 0$ we have 
\begin{equation}\label{E:max} \mathbb{E} \left[ (A_t -a)^+\right] = t -a +  a^{2} \,\mathbb{E} \left[ (a+A_t)^{-1} \exp ( \frac{ 2M_t}{a+ A_t} - \frac{2}{a} ) \right].\end{equation}
\end{theorem}
\begin{proof}Let us set $f(x,z) = x^\nu (2/y -z)$. The l.h.s. of (\ref{E:key}) becomes $E[2/y-A_t; A_t < 2/y]=E[2/y-A_t] - E[(A_t-2/y)^+]$. Since $2/y -A_t/(1+y/2\, A_t) = 1/(2/y+A_t)$, the second result comes directly from (\ref{E:key}). Using the identity (\ref{E:drift}) we can change the drift term of the exponential brownian motion to get the first result. 
\end{proof}

\vspace{1ex}

\centerline{\epsfysize=1.94in \epsfbox{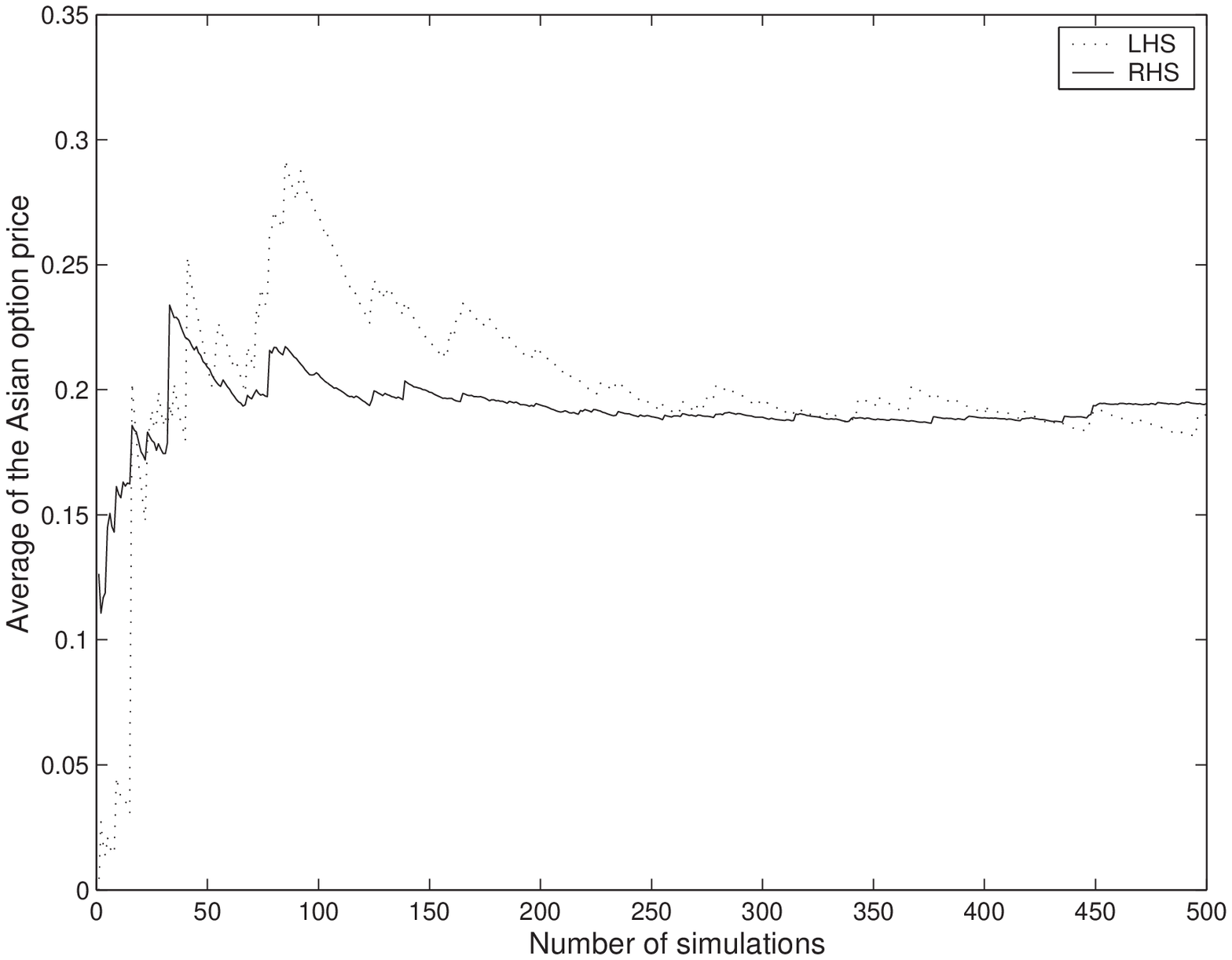}}
\centerline{Figure 2.3. Expected value of max function : $a=0.4,t=0.5$. }

\vspace{1ex}
\noindent The above figure is the simulation result of the second identity (\ref{E:max}) in Theorem \ref{T:moment}. We set $a=0.4$ and $t=0.5$. The dotted line is obtained using the l.h.s. of the equation (\ref{E:max}), that is we simulate $A_t$ using Monte Carlo methods and take the maximum function. The solid line is from the r.h.s. of the equation (\ref{E:max}). Only with 500 simulation, we can observe that the convergence is faster for the solid line, which was expected because of the discarded simulations of the l.h.s. (when $A_t$ is smaller than $a$).

Now let us consider the derivatives of $E[(A_t-a)^+]$ with respect to $a$. Since $$E[A_t -a; A_t \le a] = - \int^a_0 Pr[A_t \le u]\, du$$ we have that 
\begin{equation*}
\begin{aligned}
E[(A_t-a)^+] \ &=\ E[A_t -a] - E[A_t -a; ; A_t \le a]\\
                         &= t-a+\int^a_0 Pr[A_t \le u]\, du.
\end{aligned}
\end{equation*}
It follows that
\begin{equation}
\frac{d}{da}E[(A_t-a)^+] = -1+Pr[A_t\le a],\ \ {\rm and\ \ }\  \frac{d^2}{da^2}E[(A_t-a)^+] = g_t(a).
\end{equation}
The following theorem is the direct application of Lemma \ref{L:distribution1} and Lemma \ref{L:density}.

\begin{theorem}\label{T:derivative}
The first and second derivatives of $E[(A_t-a)^+]$ with respect to $a$ is given by following equations:
\begin{equation*}
\begin{aligned}
\frac{d}{da}E[(A_t-a)^+] \ =\ & -1+e^{-2/a} \mathbb{E} \left[  \exp \left ( \frac{2 \exp ( B_t -t/2)}{a+ A_t}\right) \right],\\
 \frac{d^2}{da^2}E[(A_t-a)^+] \ =\ &\ \frac{2}{a^2} e^{-2/a} \mathbb{E} \left[  \exp \left ( \frac{2 \exp ( B_t -t/2)}{a+ A_t}\right) \right] \\ &- \frac{1}{2}e^{-2/a} E\left[\frac{M_t}{(a/2+1/2\, A_t)^2} \exp\left( \frac{M_t}{a/2+1/2\, A_t}\right)\right].
\end{aligned}
\end{equation*}
\end{theorem}

\bigskip

\section{Sensitivities.} 
Since $A_t$ satisfies $A_{t+s} =A_t+M_t\tilde A_s$ for $t,s >0$ where $\tilde A$ is an independent copy of $A$, pricing the fixed strike Asian option at time $0 \le t \le $ the expiration date is essentially identical as the pricing the option at the beginning of the averaging period. 

One of important activities in Financial Market is managing the risk. One way to measure the risk in the option is estimating the `{\it Greek letters}' such as delta, gamma, theta, etc. In this section we use identities obtained in Section 2 to get the `{\it Greek letters}' of an European Style Asian option under the Black-Scholes framework; the price process of the risky asset follows under the risk-neutral measure
$$dS_t = \sigma S_t dB_t + r S_t dt,$$
where $B_t$ is a standard brownian motion under the risk-neutral measure, $r$ is a constant interest rate and $\sigma$ is a constant volatility of given asset. The expression (\ref{E:asian}) is equal to the price of an European style Asian option with the expiration date $\tau$ and the strike price $\kappa$ :
\begin{equation}\label{E:asiancall}
Call = \frac{S_0}{\tau \sigma^2}e^{-r\tau} E\left[( \int_0^{\sigma^2\tau}  e^{ B_t - t/2}\, dt - \frac{\sigma^2\kappa\tau}{S_0})^+\right ].
\end{equation}

\begin{theorem}\label{T:call}
The price of an Asian call at time $t=0$ is given by
\begin{equation*}
Call_0 = S_0 e^{-r \tau} \left[ 1- \frac{\kappa}{S_0} + \frac{\sigma^2 \kappa^2 \tau}{S_0^2} E \left[(a+A_{\tau \sigma^2})^{-1} \exp \left( \frac{2 M_{\tau\sigma^2}}{a+A_{\tau\sigma^2}}-\frac{2}{a}\right)\right]\right],
\end{equation*}
where $a= \frac{\sigma^2\kappa\tau}{S_0}$, $\tau$ is the expiration date, $\kappa$ is the strike price.
\end{theorem}

\begin{proof} The result comes directly from the combination of Theorem \ref{T:moment} and equation (\ref{E:asiancall}).\end{proof}

The delta $\Delta$ is the rate of change of the price of the option with respect to the price of the underline asset. Thus we have 
\begin{equation}\label{E:delta}\Delta =\frac{ Call }{S_0} -\left. \frac{\kappa}{S_0} e^{-r \tau} \frac{d}{da} E[(A_{\sigma^2 \tau} -a)^+]\right |_{a= \frac{\sigma^2\kappa\tau}{S_0}}.\end{equation}
Using Theorem \ref{T:moment} and Theorem \ref{T:derivative} we have 

\begin{theorem}\label{T:delta}
The delta $\Delta_0$ of an Asian call option at time $t=0$ is given by
\begin{equation*}
\begin{aligned}
\Delta_0= e^{-r\tau} + \frac{\kappa}{S_0} e^{-r \tau-2/a}& \left\{ a E \left[(a+A_{\tau \sigma^2})^{-1} \exp \left( \frac{2 M_{\tau\sigma^2}}{a+A_{\tau\sigma^2}}\right)\right]\right.\\&\left.\left.- E\left[ \exp \left( \frac{2 M_{\tau\sigma^2}}{a+A_{\tau\sigma^2}}\right)\right]\right\}\right|_{a= \frac{\sigma^2\kappa\tau}{S_0}}.
\end{aligned}
\end{equation*}
\end{theorem}

The gamma $\Gamma$ is the rate of change of the delta with respect to the price of underlying asset. Thus the direct computation using equation (\ref{E:delta}) and the results in Theorem \ref{T:moment}, Theorem \ref{T:derivative} and Theorem \ref{T:delta} we have 
\begin{theorem}\label{T:gamma}
The gamma $\Gamma_0$ of an Asian call option at time $t=0$ is given by
\begin{equation*}
\Gamma_0= \frac{\sigma^2 \kappa^2\tau}{S_0^3} e^{-r \tau}  g_{\tau\sigma^2} (\frac{\sigma^2\kappa\tau}{S_0}),\end{equation*}
where $g_t$ is a continuous probability density function of $A_t$ given in Lemma \ref{L:density}.
\end{theorem}

\vspace{1ex}

\centerline{\epsfysize=1.4in \epsfbox{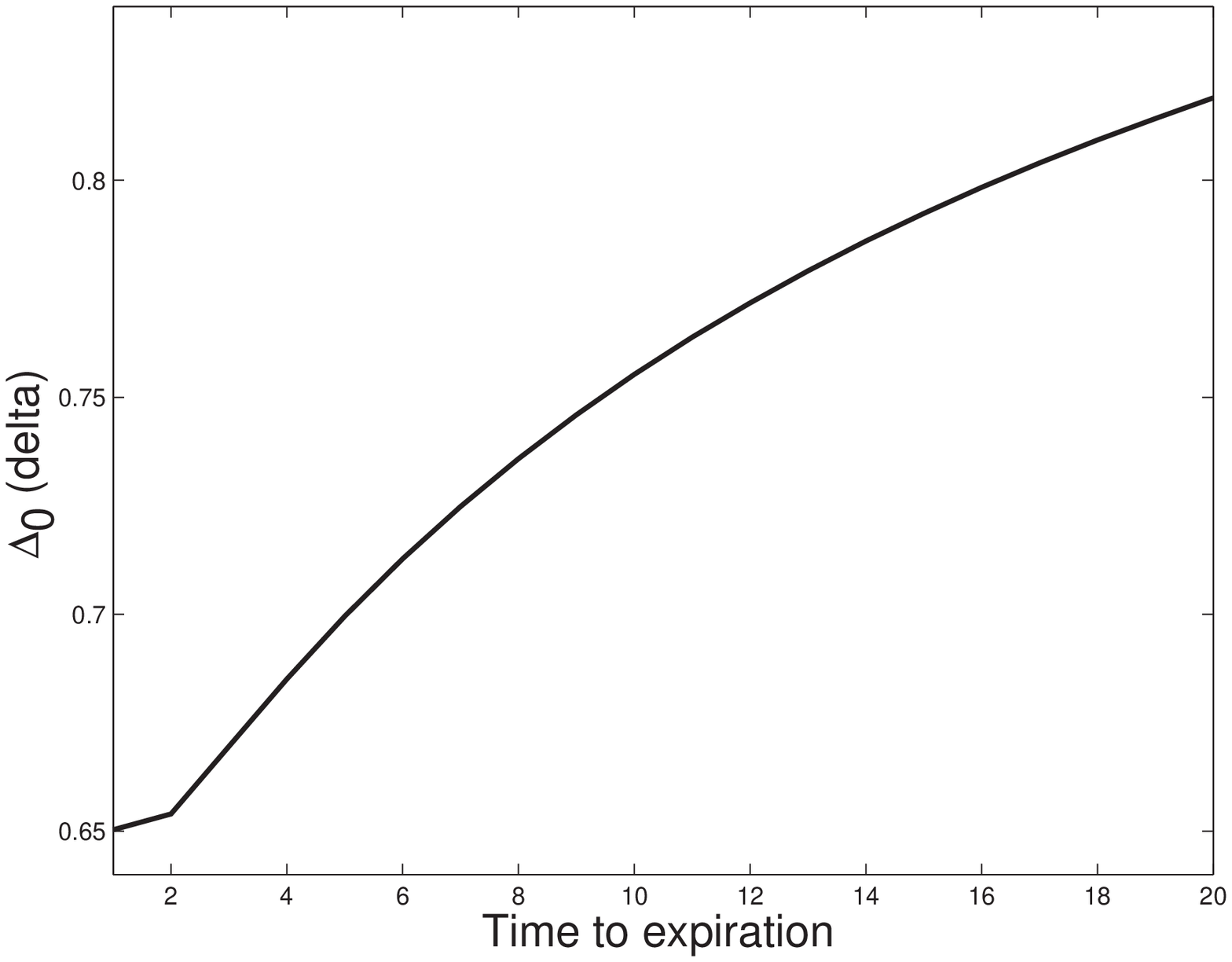}\quad \epsfysize=1.4in \epsfbox{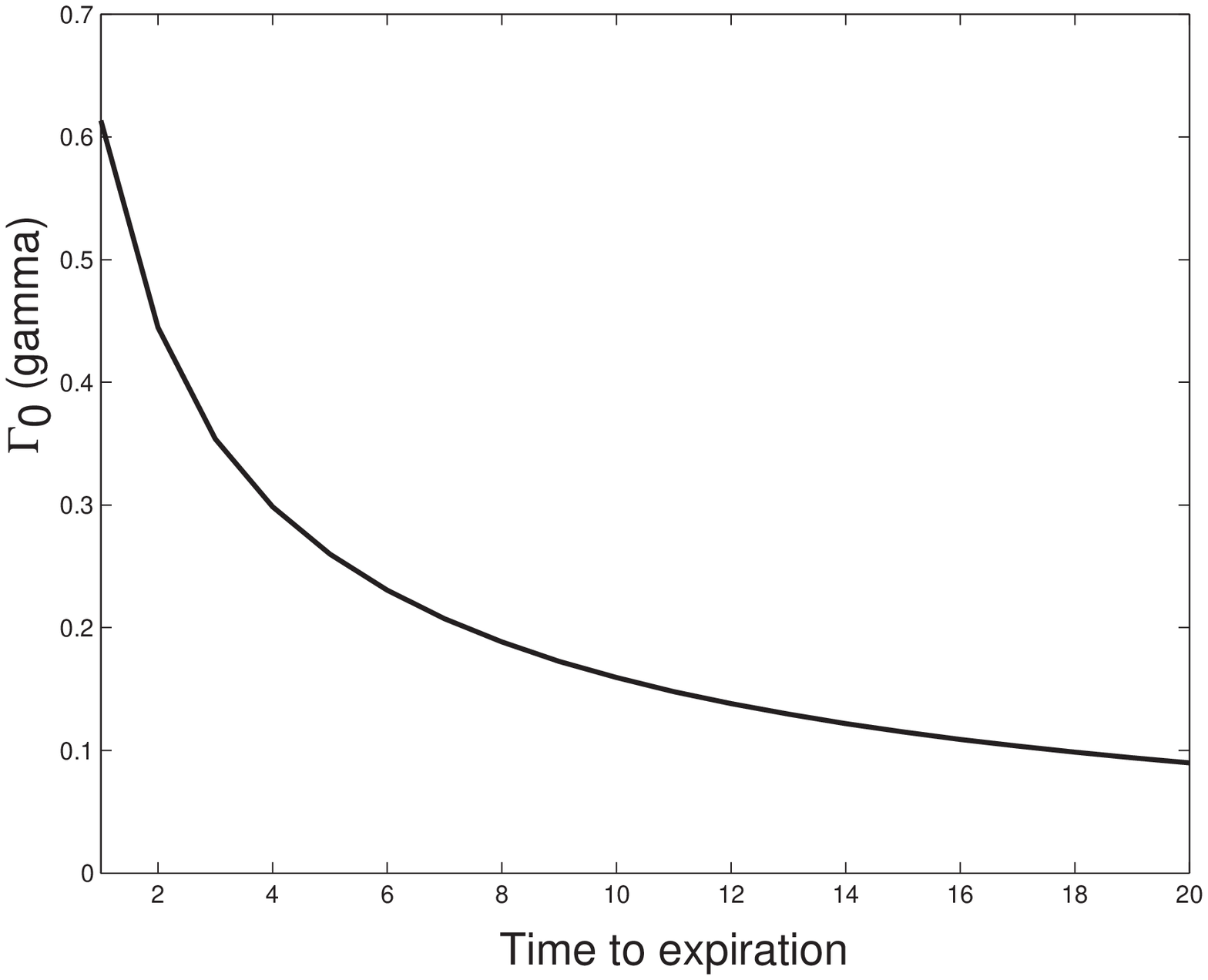}}
\centerline{\epsfysize=1.4in \epsfbox{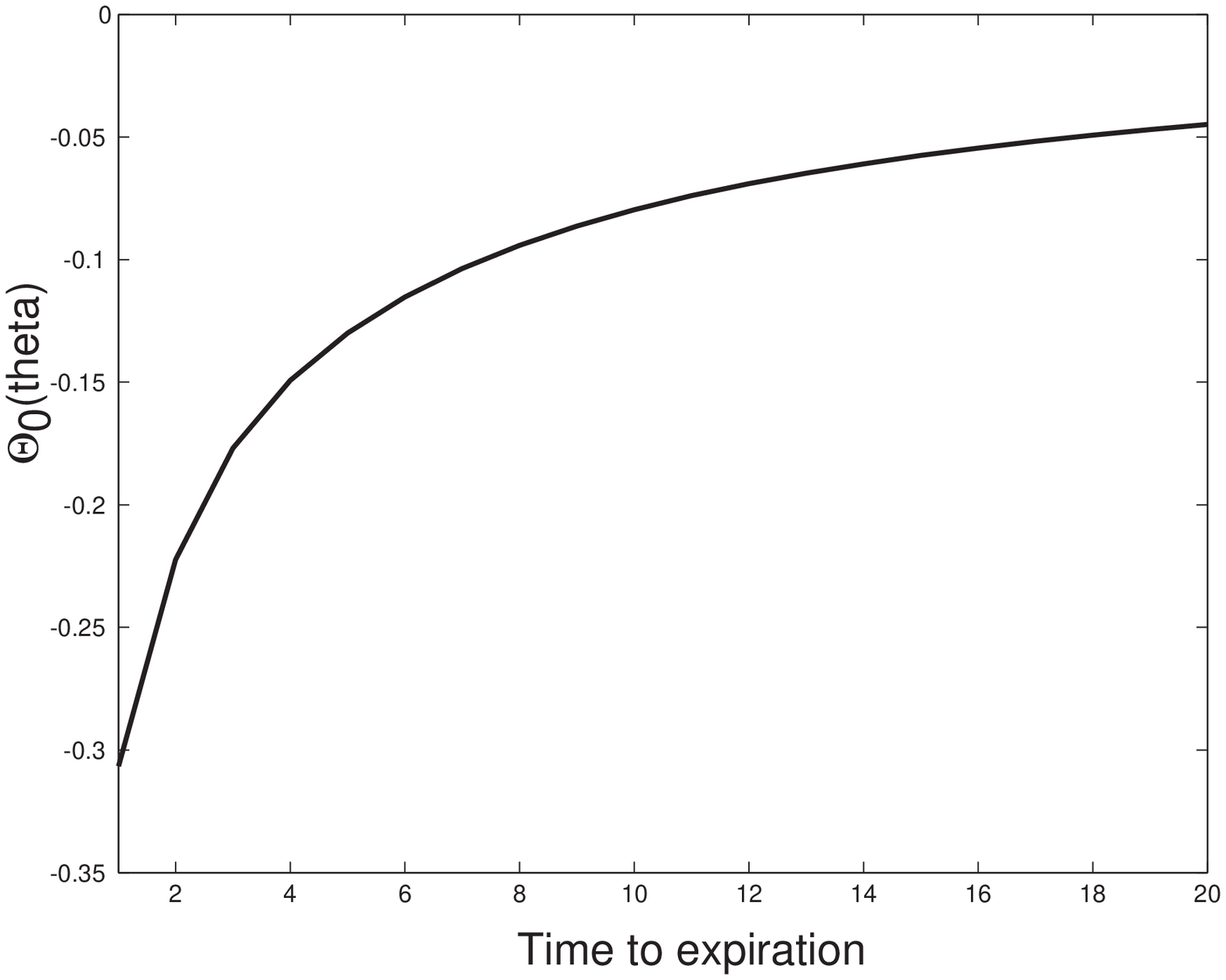}\quad \epsfysize=1.4in \epsfbox{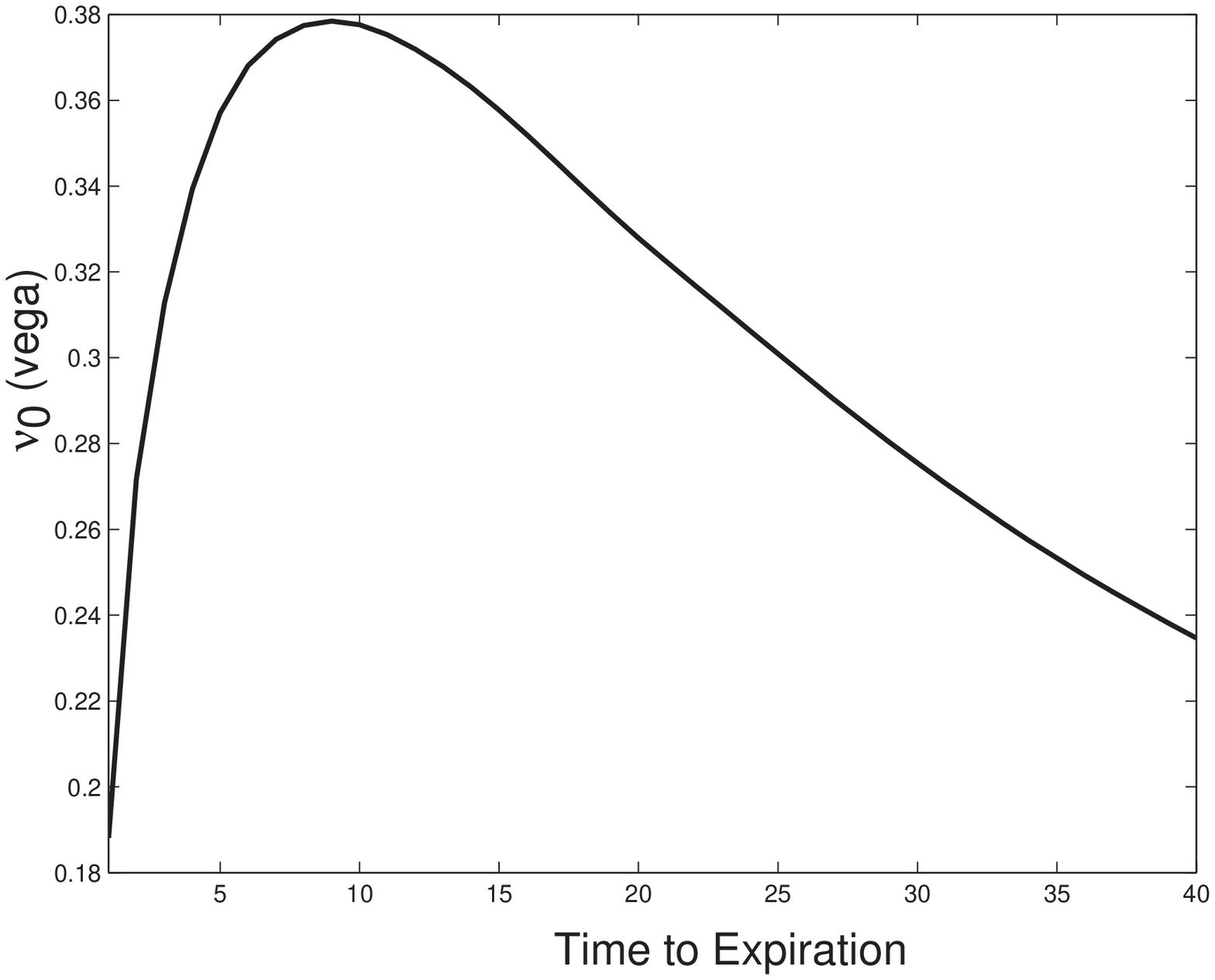}}
\centerline{Figure 3.1. Derivatives vs. the time to expiration: $S_0=\kappa=\sigma=1$ }

\vspace{1ex}
In Figure 3.1. we use Monte Carlo methods to get Greek letters $\Delta, \Gamma$, $\Theta$ and $\nu$. We take the risk free rate $r=0$ and both the initial stock price and the strike price are equal to 1. We also set for simplicity $\sigma=1$. Notice that we plot greek letters vs. time to expiration. Since we computed greeks at the beginning of the averaging period, time to expiration is the same as the length of the averaging period. In the upper l.h.s. figure we can observe that as time to expiration increases, the value of $\Delta_0$ increases. The plot of $\Gamma_0$ with respect to time to expiration is given in the upper r.h.s. figure.  Since $\Gamma$ is the first derivative of $\Delta$ with respect to the stock prices, it reflects our observation in the plot of $\Delta_0$. The lower l.h.s. figure is the plot of $\Theta_0$ and the lower r.h.s. figure is the plot of $\nu_0$.

Since all parameters are constants, the call price satisfies the differential equation
$$\frac{\partial Call}{\partial t} + r S \frac{\partial Call}{\partial S} + \frac{1}{2} \sigma^2 S \frac{\partial^2 Call}{\partial S^2} = r \,Call.$$
It follows that 
$$\Theta_0 + r S_0 \Delta_0 + \frac{1}{2} \sigma^2 S_0 \Gamma_0 = r \,Call_0$$
where $\Theta_0$ is the rate of change of the price of the option at time $t=0$ with respect to time and $Call_0$ is the price of the call option at time $0$.

\begin{theorem}\label{T:theta}
The theta $\Theta_0$ of an Asian call option at time $t=0$ is given by
\begin{equation*}
\Theta_0= r Call_0 -r S_0 \Delta_0 - \frac{1}{2} \sigma^2 S_0 \Gamma_0
\end{equation*}
where $Call_0$, $\Delta_0$ and $\Gamma_0$ are given in the previous theorems 3.1--3.
\end{theorem}

Now let us discuss the vega ${\bf \nu}$. The vega $\mathbf{\nu}$ is the rate of change of the price of the option with respect to the volatility of the underlying asset. The call price of European style Asian option is given in (\ref{E:asiancall}). Thus we have

\begin{equation*}
\begin{aligned}
\frac{\partial Call }{\partial \sigma} &= -\frac{2}{\sigma} \, Call + \frac{2S_0}{\sigma}e^{-r \tau} E\left[ M_{\sigma^2 \tau}- \frac{\kappa}{S_0} \ ;\ A_{\sigma^2 \tau} > \frac{\sigma^2 \kappa\tau}{S_0} \right]\\
&= -\frac{2}{\sigma} \, Call +\frac{2S_0}{\sigma} e^{-r \tau} E\left[M_{\sigma^2 \tau}- \frac{\kappa}{S_0}\right] \\&\phantom{alsdkj} - \frac{2S_0}{\sigma} e^{-r \tau} E\left[M_{\sigma^2 \tau}- \frac{\kappa}{S_0}\,;\, A_{\sigma^2 \tau} < \frac{\sigma^2 \kappa\tau}{S_0} \right].
\end{aligned}
\end{equation*}
It follows that 
\begin{equation}\label{E:vega}
\begin{aligned}
\frac{\partial Call }{\partial \sigma} &= -\frac{2}{\sigma} \, Call+ \frac{2S_0}{\sigma} e^{-r \tau} \left( 1- E\left[M_{\sigma^2 \tau} \,;\, A_{\sigma^2 \tau} < \frac{\sigma^2 \kappa\tau}{S_0} \right]\right)\\&-\frac{2 \kappa}{\sigma} \left(1-P \left[A_{\sigma^2 \tau}<  \frac{\sigma^2 \kappa\tau}{S_0} \right]\right).
\end{aligned}
\end{equation}
By setting $\nu=1$ in equation (\ref{E:drift}) in Lemma \ref{L:distribution1}, we see that 
$$1- E\left[M_{\sigma^2 \tau}\,;\, A_{\sigma^2 \tau} < \frac{\sigma^2 \kappa\tau}{S_0} \right] = P \left[A_{\sigma^2\tau}^{(1)}> \frac{\sigma^2 \kappa\tau}{S_0} \right]. $$
Therefore we have 
\begin{theorem}\label{T:vega}
The vega $\nu_0$ of an asian call option at time $t=0$ is given by 
\begin{equation*}
\nu_0 = -\frac{2}{\sigma} \, Call+  \frac{2S_0}{\sigma} e^{-r \tau} P \left[A_{\sigma^2\tau}^{(1)}> \frac{\sigma^2 \kappa\tau}{S_0} \right]-\frac{2 \kappa}{\sigma}P \left[A_{\sigma^2 \tau}>  \frac{\sigma^2 \kappa\tau}{S_0} \right].
\end{equation*}
\end{theorem}

\noindent Last two quantities in the above theorem can be obtained using Lemma \ref{L:distribution2}.

\end{document}